\documentclass[usenatbib]{mnras}
\usepackage{dcolumn}
\usepackage{graphicx}
\usepackage{url}
\usepackage{color}
\usepackage[T1]{fontenc}
\usepackage{ae,aecompl}
\usepackage{pdflscape}

\newcommand{\cm}{cm$^{-1}$}

\newcommand{\ai}{\textit{ab initio}}

\newcommand{\etal}{\textit{et al.}}

\newcommand{\eqref}[1]{(\ref{#1})}

\newcommand{\sotwo}{SO$_2$}

\newcommand{\pp}{^{\prime\prime}}

\title[ExoMol XIV: Line list for SO$_2$]{ExoMol molecular line lists  - XIV: The rotation-vibration spectrum of hot SO$_2$}
\date{\today}
\author[D.S. Underwood et al.]{Daniel S. Underwood$^{1}$,  Jonathan Tennyson$^{1}$\thanks{Email: j.tennyson@ucl.ac.uk}, Sergei N. Yurchenko$^{1}$,
\newauthor Xinchuan Huang$^{2}$, David W. Schwenke$^{3}$, Timothy J. Lee$^{4}$,
\newauthor S{\o}nnik Clausen$^5$ and Alexander Fateev$^{5}$
\\
$^{1}$Department of Physics and Astronomy, University College London, London WC1E 6BT, UK\\
$^{2}$SETI Institute, Mountain View, CA 94043 USA\\
$^{3}$NASA Ames Research Center, NAS Facility, Moffett Field, CA 94035 USA\\
$^{4}$NASA Ames Research Center, Space Science \&\ Astrobiology Division, Moffett Field, CA 94035 USA\\
$^{5}$Technical University of Denmark, Department of Chemical and Biochemical Engineering, Frederiksborgvej 399,
4000 Roskilde, Denmark}
\date{Accepted XXXX. Received XXXX; in original form XXXX}
\pagerange{\pageref{firstpage}--\pageref{lastpage}}
\pubyear{2016}

\begin{document}
\label{firstpage}

\maketitle

\begin{abstract}
  Sulphur dioxide is well-known in the atmospheres of planets and
  satellites, where its presence is often associated with volcanism,
  and in circumstellar envelopes of young and evolved stars as well as
  the interstellar medium. This work presents a line list of 1.3 billion
  $^{32}$S$^{16}$O$_2$ vibration-rotation transitions computed using
  an empirically-adjusted potential energy surface and an {\it ab
    initio} dipole moment surface. The list gives complete
  coverage up to 8000 \cm\ (wavelengths longer than 1.25 $\mu$m) for
  temperatures below 2000 K. Infrared absorption cross sections are
  recorded at 300 and 500 C are used to validated the resulting
  ExoAmes line list.  The line list is made available in
  electronic form as supplementary data to this article and at
  \url{www.exomol.com}.

\end{abstract}


\section{Introduction}

Suphur dioxide, SO$_2$,  has been
 detected
in a variety of astrophysical settings. Within the solar
system, SO$_2$ is known to be a major  constituent of the
atmospheres of Venus
\citep{79Barker.SO2,12BeMoBe.SO,08BeKoFe.SO2,14ArMeCr.SO2} and Jupiter's moon, Io
\citep{79PeHaMa.SO2,80NeLaMa.SO2,94BaMcSt.SO2}.  SO$_2$ has been observed in the atmosphere of Mars,
although to a much lesser extent \citep{15KhViMu}. Volcanic activity is an important indicator
of the presence of SO$_2$.

The chemistry of sulphur-bearing
species, including SO$_2$,  has been studied in the
atmospheres of giant planets, brown dwarfs, and dwarf stars by \citet{06ViLoFe.SO2}.
SO$_2$ has been observed in
circumstellar envelopes of young and evolved stars
\citep{99YaJoOs.SO2,03TaBoBr.SO2,06Zixxxx.SO2,13AdEdZi.SO2}, and in
molecular clouds and nebulae within the interstellar medium
\citep{97KlScBe.SO2,01ScBeHu.SO2,10CrBeWa.SO2,13BeMuMe.SO2}.
Extragalactic detection of SO$_2$ has even been achieved
\citep{03MaMaMa.SO2,05MaMaMa.SO2}, emphasising the universal abundance of this
particular molecule.

SO$_2$ is known to occur naturally in Earth's atmosphere where it is
found in volcanic emissions and hot springs
\citep{73StJexx.SO2,15KhViMu.SO2} where observation of gases such as
SO$_2$ provide a useful tool in the understanding of such geological
processes. The spectroscopic study of SO$_2$ can also provide insight
into the history of the Earth's atmosphere \citep{13WhXiHu.SO2}.
However its most important impact is arguably through its contribution
to the formation of acid rain \citep{14HiMexx.SO2} where the
oxidisation of SO$_2$ to SO$_3$ in the atmosphere, followed by
subsequent rapid reaction with water vapour results in the production
of sulphuric acid (H$_2$SO$_4$), which leads to many adverse
environmental effects.  Spectra of hot SO$_2$ are also important for
technological applications such as monitoring engine exhausts
\citep{13VoKaEg.SO2}, combustion \citep{14HiMexx.SO2} and etching
plasmas \citep{90GrHaxx.SO}.

One of the most exciting astronomical developments in recent years is
the discovery of extrasolar planets, or ``exoplanets''. The
observation of the tremendous variety of such bodies has challenged
the current understanding of solar system and planetary formation.
Exoplanet detection methods have grown in sophistication since the
inception of the field, however efforts to characterise their
atmospheres are relatively new \citep{13TiEnCo.exo}. The
well-documented distribution of sulphur oxides in various terrestrial
and astrophysical environments means that a thorough understanding of
their fundamental spectroscopic behaviour is essential in the future
analysis of the spectra of these exoplanetary atmospheres, and of
other bodies of interest observed through past, present and future
space telescope missions \citep{14HuScLe.SO2,13KaLoDo.SO2}.

Experimentally, SO$_2$ spectra have been studied in both the
ultra-violet
\citep{84FrYoEs.SO2,99StSmRu.SO2,03RuStSm.SO2,08DaEsJo.SO2,09Lyonsx.SO2,09RuStSm.SO2,11BlBlSt.SO2,12DaHaSh.SO2,13FrDaFa.SO2,15EnDaUe}
and infrared
\citep{92LaFrPi.SO2,93FlLaxx.SO2,93LaPiFl.SO2,96LaPiHi.SO2,98ChWeLa.SO2,08HeBaBa.SO2,09UlBeHo.SO2,10UlBeGr.SO2,11UlGrBe.SO2,12UlGrBe.SO2,13UlOnGr.SO2}
at room-temperature; most of these data are captured in the HITRAN
database \citep{jt557}.  Conversely, there is limited spectral data
for SO$_2$ available at elevated temperatures, and much of it is
either not applicable to the spectral region of interest or consists
of remote observational data requiring a sophisticated, bespoke
atmospheric model to be used in conjunction with a line list to reproduce it
\citep{13GrFaNi.SO2,15KhViMu}. However a few measurements of cross
section data have been made for hot SO$_2$ spectra in the laboratory
by \citet{13GrFaNi.SO2} and \citet{14GrFaCl.SO3}.  Here we extend this
work by recording spectra of SO$_2$ in the infrared as a function of
temperature for comparison with and validation of our computed line
list.

Theoretically a number of studies have looked at the ultraviolet spectrum
spectrum of SO$_2$ \citep{00XiGuBl.SO2,07RaXiGu.SO2,15LeTaKo.SO2} which
represents a considerable challenge. More straightforward are studies
of the vibration-rotation spectrum which lies in the infrared. Early
work on this problem was performed by
\citet{92KaHaxx.SO2} while recent work has focused on
rotational excitation \citet{15KuPoxa.SO2,15KuPoxb.SO2}. A number
of comprehensive studies has been performed by the Ames group
\citep{14HuScLe.SO2,15HuScLe.SO2,16HuScLe.SO2}; this work
provides an important precursor to this study and will be discussed
further below.

The ExoMol project \citep{jt528} aims to provide molecular line lists
for exoplanet and other atmospheres with a particular emphasis on
hot species.  \citet{14HuScLe.SO2} used theoretical methods
to compute a line list for \sotwo\ up to 8000 \cm\ for a temperature
of 296 K (denoted Ames-296K). This was recently extended to 5
isotopologues of SO$_2$ \citep{15HuScLe.SO2}.  This work and
methodology follow closely similar studies  on H$_2$O
\citep{97PaScxx.H2O}, NH$_3$ \citep{08HuScLe,11HuScLe.NH3,11HuScLe2.NH3}, and
CO$_2$ \citep{12HuScTa.CO2}. In this work we build on the work of
\citet{14HuScLe.SO2} to compute a line list for hot SO$_2$ which
should be appropriate  for temperatures approaching
2000~K. Doing this required some technical adjustments both to the
potential energy surface (PES) used and nuclear motion program
employed; these are described in section 2. Section 3 presents our
experimental work and section 4 the line list computations. Results
and comparisons are given in section 5. Section 6 gives our
conclusions.

\section{Theoretical method}

In order to compute a line list for \sotwo\ three things are required:
a suitable potential energy surface (PES), dipole moment surface
(DMS), and a nuclear motion program \citep{jt475}. The DMS used here
is the \ai\ one of \citet{14HuScLe.SO2} and is based on 3638
CCSD(T)/aug-cc-pV(Q+d)Z level calculations. The other parts are
consider in the following subsections.

\subsection{Potential Energy Surface}

The Ames-1B PES used here is spectroscopically determined by refining
an {\it ab initio} PES using room-temperature spectroscopic data.  The
Ames-1B PES refinement procedure used is very similar to the Ames-1
refinement reported by \citet{14HuScLe.SO2}. The two main differences
are the choice of the initial PES and the use of a now-converged
stretching basis. The published Ames-1 PES was chosen as the initial
PES to adjust. All 22 zeroth- to fourth-order coefficients of the
short-range PES terms are allowed to vary although the zeroeth-order
constant does not affect the results. The number of reliable HITRAN
\citep{jt557} energy levels included with $J=0/4/20/50/70$ are
23/43/183/181/129, respectively.  The corresponding weights adopted
for most levels are 2.5/1.0/1.5/2.0/3.0, respectively.  For the
original Ames-1 PES coefficients, the initial root mean square fit
error, $\sigma_{\rm rms}$, are 0.175 \cm\ (weighted) and 0.085 \cm\
(unweighted).  The refined coefficient set significantly reduces
$\sigma_{\rm rms}$ to 0.028 \cm\ (weighted) and 0.012 \cm\
(unweighted). The PES is expressed in changes from equilibrium values
of the bond lengths ($\Delta r_1, \Delta r_2$) and bond angle
($\Delta\alpha$).  Compared to the Ames-1 PES coefficients, the
largest percentage variations are $\pm 11-22$\%\ found for the
following short-term expansion terms: $\Delta\alpha$ gradient, $\Delta
r$ gradient, $\Delta r(\Delta\alpha)^3$, $\Delta r_1 \Delta r_2$,
$(\Delta r)^2(\Delta\alpha)^2$ and $(\Delta r_1)^3 \Delta r_2$, The
changes in absolute value are largest for $(\Delta r_1)^2 \Delta\alpha
\Delta r_2$ and $(\Delta r_1)^2 (\Delta r_2)^2$. The Ames-1B PES has
been used in recent SO$_2$ isotopologue calculations
\citep{16HuScLe.SO2} and is available upon request.

The accuracy of the mass-independent Ames-1B PES remains approximately
the same as the Ames-1 PES: about 0.01 \cm\ for the three vibrational
fundamentals of the three main isotopologues,
646, 828 and 628 in HITRAN notation, which therefore we can expect
similar accuracy for the fundamentals of
the minor isotopologues 636, 627 and 727 \citep{16HuScLe.SO2}.
For vibrational states as high as 5500 \cm, e.g. $5\nu_1$,
accuracy for isotopologues using the Ames-1B PES
should be better than 0.05 \cm. For those
energy levels far beyond the upper limit included in our empirical
refinement, e.g. 8000 -- 10,000 \cm\ and above, the accuracy would
gradually degrade to a few wavenumbers, approximately the quality of
the original {\it ab initio} PES  before empirical refinement. The
agreement between VTET and DVR3D results is better than 0.01 \cm\ up
to at least 8000 \cm. However, it should be noted that the Ames-1B
PES was refined using the VTET program in such a way that although
less than 500 HITRAN levels were adopted in the refinement, the
accuracy is consistently as good as 0.01 -- 0.02 \cm\ from 0 to 5000
\cm. The accuracy mainly depends on the energy range covered by the
refinement dataset, but not on whether a particular energy level was
included in the refinement. Recent experiments have verified the
prediction accuracy for non-HITRAN energy levels and bands. New
experimental data at a higher energy range may significantly extend
the wavenumber range with an 0.01 -- 0.02 \cm\ accuracy, provided a new
refinement is performed with the new data. Currently, the accuracy of the
reported line list is best at 296~K and below 7000 \cm. Higher
energy levels may have errors ranging from 0.5 \cm\
to a few \cm.

Use of the Ames-1B PES with the VTET nuclear motion program
\citep{96Schwen.method} employed by the Ames group and DVR3D
\citep{jt338} employed here gives very similar results for room
temperature spectra. However, one further adjustment was required for
high $J$ calculations as the PES appears to become negative at very
small ${\rm H\hat{S}H}$ angles. Similar problems have been encountered
before with water potentials \citep{92ChLixx.H2O,97PaScxx.H2O,jt308}
which have been overcome by adding a repulsive H -- H term to the PES.
Here we used a slightly different approach. The bisector-embedding
implementation in DVR3D has the facility to omit low-angle points from
the calculation \citep{jt114}; usually only a few automatically-chosen
points are omitted.  An amendment to module DVR3DRJZ allowed for the
selection of the appropriate PES region by omitting all low-angle
functions beyond a user-specified DVR grid point. This amendment was
essential for the high $J$ calculations.  This version of the code was
used for all calculations with $J \geq$ 50 pressented. For $J <$ 50
this defect only affected very high rovibrational energies computed in
ROTLEV3B and there has no significant effect on the results.

\subsection{Nuclear motion calculations}

The line list was produced using the
DVR3D program suite \citep{jt338} and involved
rotationally excited states up to $J=165$.
As this  doubled the highest $J$ value  previously computed using
DVR3D, a number of adjustments were necessary compared to the
published version of the code.

Firstly, the improved rotational Hamiltonian construction algorithm
implemented by \citet{jtH2S} was employed; this proved vital to making
the calculations tractable. Secondly, it was necessary to adjust the
automated algorithm which generates (associated) Gauss-Legendre
quadarature points and weigths: the previous algorithm failed for
grids of more than 90 points but a simple brute-force search for
zeroes in the associated Legendre $P_N^k(x)$ was found to work well
for all $N$ and $k$ values tested ($N \leq 150$).  Thirdly, the DVR3D
algorithm relies on solving a Coriolis-decoupled vibrational problem
for each $(J,k)$, where $k$ is the projection of the rotational motion
quantum number, $J$, onto the chosen body-fixed $z$ axis. This
provides a basis set from which functions used to solve the
fully-coupled rovibrational problem are selected on energy grounds \cite{jt66}.
Our calculations
showed that for $k$ values above 130 no functions were selected so an
option was implemented in which only $(J,k)$ combinations which
contributed functions were actually considered. This update does not
save significant computer time, since the initial $(J,k)$ calculations
are quick, but does reduce disk usage which also proved to be a
significant issue in the calculations. Finally, it was found that
algorithm used in module DIPOLE3 to read in the wavefunctions led to
significant input/output problems. This module was re-written to
reduce the number of times these wavefunctions needed to be read.  The
updated version of the DVR3D suite is available in the DVR3D project
section of the CCPForge program depository
(https://ccpforge.cse.rl.ac.uk/).


DVR3D calculations were performed in Radau coordinates using the so-called
bisector embedding \citep{jt114} which places the body-fixed axis close
to standard A principle axis of SO$_2$ meaning the $k$ used by the
program is close to the standard asymmetric top approximate quantum number
$K_a$. The DVR (discrete variable representation) calculations are
based on grid points corresponding to Morse oscillator-like functions
\citep{jt14} for the stretches and (associated) Legendre polynomials
for the bend.

The rotational step in the calculation selects
the lowest  $n(J + 1)$ functions from the $(J,k)$-dependent vibrational
functions on energy grounds where $n$ is parameter which was chosen
to converge the rotational energy levels. For $J < 100$ it was found
that $n=725$ was required to get good convergence. However such calculations
become computationally expensive for high $J$ values for which, in practice,
fewer levels are required.
Reducing the value of $n$  to 500 was found to
have essentially no effect on the convergence of rovibrational
eigenvalues produced at $J$ = 124. For $J=124$, there are a total
of 14~523 eigenvalues below 15~000 \cm\ summed over all rotational symmetries.
This constitutes roughly 8\%\ of the total combined matrix
dimension of 175~450 for $n$ = 725 and 12\%\ of 121 000 for $n$ = 500.
The value $n$ = 725 was originally obtained for convergence of
energies at $J$ = 60, where the number of eigenvalues below 15 000
\cm\ accounts for 38\%\ of the combined matrix dimension of 88~450.
The higher energies at $J$ = 60 are much more sensitive to the value
of $n$ due to the way the basis functions are
distributed, whereas for $J \geq$ 120 the energies below the 15 000
\cm\ threshold are already easily converged at lower values of $n$.
It was therefore decided to reduce
$n$ to 500 for $J \geq 124$ which leads to
minimal loss of accuracy.

Convergence of rovibrational energy levels with $n = 725$ was obtained
using $J = 60$ calculations. The sum of energies below 10 000, 11 000,
12 000, 13 000, 14 000, and 15000 \cm\ was used to give an indication
of the convergence below those levels (see Table 3.1 in
\citet{16Underw.SO2} Thesis]). $J = 60$ coincides with the largest
number of ro-vibrational energies lying below 15 000 \cm, and thus the
higher energies here are the most sensitive to the convergence tests.
The value $n = 725$ ensures that the sum of all energies below 10 000
\cm\ and 15 000 \cm\ are fully converged to within 0.0001 \cm\ and to
1 \cm, respectively. Computed DVR3D rovibrational energies above 10
000 \cm, though converged, can not be guaranteed to be
spectroscopically accurate, and are entirely dependent on the quality
of the PES. This may have minor repercussions on the convergence
of the partition function at the high end of the temperature range.
Table~\ref{tab:dvr3d} summarises the parameters used in the DVR3D
calculations while Table~\ref{AMES-1B-agree-even} compares vibrational
term values computed with DVR3D and VTET.
\begin{table*}
\caption{Input parameters for DVR3DRJZ and ROTLEV3B modules of DVR3D \citep{jt338}.}\label{tab:dvr3d}
\small
\begin{tabular}{lll}
\hline\hline
 \multicolumn{3}{l}{DVR3DRJZ}\\
\hline
Parameter & Value & Description\\
\hline
NPNT2 & 30 & No. of radial  DVR points (Gauss-Laguerre)\\
NALF & 130 & No. of angular DVR points (Gauss-Legendre)\\
NEVAL &1000 & No. of eigenvalues/eigenvectors required\\
MAX3D & 3052 & Dimension of final vibrational Hamiltonian\\
XMASS (S) & 31.963294 Da & Mass of sulphur atom\\
XMASS (O) & 15.990526 Da& Mass of oxygen atom\\
$r_e$ & 3.0~a$_0$ & Morse parameter (radial basis funciton)\\
$D_e$ & 0.4~E$_h$ & Morse parameter (radial basis funciton)\\
$\omega_e$ & 0.005~a.u. & Morse parameter (radial basis funciton)\\
\hline
 \multicolumn{3}{l}{ROTLEV3B}\\
\hline
Parameter & Value & Description\\
\hline
NVIB & 1000 & No. of vib. functions read fot  each
$k$\\
$n$ & 725 & Defines $IBASS = n(J + 1)$ for $J <$ 124\\
$n$ & 500 & Defines $IBASS = n(J + 1)$ for $J \geq$ 124\\
\hline
\hline
\end{tabular}
\end{table*}

\begin{table}
\caption{A comparison of even symmetry vibrational bands in \cm\ based on the AMES-1B PES.}
\centering
\begin{tabular}{lrrr}
\hline\hline
Band & VTET  & DVR3D & Difference \\
\hline

$\nu_2$	&	517.8725	&	517.8726	&	-0.0001	\\
$2\nu_2$	&	1035.1186	&	1035.1188	&	-0.0002	\\
$\nu_1$	&	1151.7138	&	1151.7143	&	-0.0005	\\
$3\nu_2$	&	1551.7595	&	1551.7598	&	-0.0003	\\
$\nu_1 + \nu_2$	&	1666.3284	&	1666.3288	&	-0.0004	\\
$4\nu_2$	&	2067.8084	&	2067.8087	&	-0.0003	\\
$\nu_1 + 2\nu_2$	&	2180.3187	&	2180.3191	&	-0.0004	\\
$2\nu_1$	&	2295.8152	&	2295.8158	&	-0.0006	\\
$5\nu_2$	&	2583.2704	&	2583.2708	&	-0.0004	\\
$\nu_1 + 3\nu_2$	&	2693.7053	&	2693.7056	&	-0.0003	\\
$2\nu_3$	&	2713.3936	&	2713.3938	&	-0.0002	\\
$2\nu_1 + \nu_2$	&	2807.1739	&	2807.1744	&	-0.0005	\\
$6\nu_2$	&	3098.1428	&	3098.1432	&	-0.0004	\\
$\nu_1 + 4\nu_2$	&	3206.5009	&	3206.5012	&	-0.0003	\\
$\nu_2 + 2\nu_3$	&	3222.9523	&	3222.9526	&	-0.0003	\\
$2\nu_1 + 2\nu_2$	&	3317.9078	&	3317.9082	&	-0.0004	\\
$3\nu_1$	&	3432.2724	&	3432.2729	&	-0.0005	\\
$7\nu_2$	&	3612.4145	&	3612.4150	&	-0.0005	\\
$\nu_1 + 5\nu_2$	&	3718.7109	&	3718.7111	&	-0.0002	\\
$2\nu_2 + 2\nu_3$	&	3731.9370	&	3731.9373	&	-0.0003	\\
$2\nu_2 + 3\nu_2$	&	3828.0367	&	3828.0370	&	-0.0003	\\
$\nu_1 + 2\nu_3$	&	3837.6154	&	3837.6161	&	-0.0007	\\
$3\nu_1 + \nu_2$	&	3940.3781	&	3940.3786	&	-0.0005	\\
$8\nu_2$	&	4126.0668	&	4126.0673	&	-0.0005	\\
$\nu_1 + 6\nu_2$	&	4230.3325	&	4230.3327	&	-0.0002	\\
$3\nu_2 + 2\nu_3$	&	4240.3549	&	4240.3553	&	-0.0004	\\
$2\nu_1 + 4\nu_2$	&	4337.5726	&	4337.5729	&	-0.0003	\\
$\nu_1 + \nu_2 + 2\nu_3$	&	4343.8153	&	4343.8158	&	-0.0005	\\
$3\nu_1 + 2\nu_2$ 	&	4447.8567	&	4447.8572	&	-0.0005	\\
$4\nu_1$	&	4561.0634	&	4561.0638	&	-0.0004	\\
$9\nu_2$	&	4639.0726	&	4639.0731	&	-0.0005	\\
$\nu_1 + 7\nu_2$	&	4741.3554	&	4741.3556	&	-0.0002	\\
$4\nu_2 + 2\nu_3$	&	4748.2069	&	4748.2074	&	-0.0005	\\
$2\nu_1 + 5\nu_2$	&	4846.5190	&	4846.5192	&	-0.0002	\\
$\nu_1 + 2\nu_2 +2\nu_3$	&	4849.4433	&	4849.4438	&	-0.0005	\\
$2\nu_1 + 2\nu_3$	&	4953.5971	&	4953.5978	&	-0.0007	\\
$3\nu_1 + 3\nu_2$	&	4954.7285	&	4954.7289	&	-0.0004	\\
$4\nu_1 + \nu_2$	&	5065.9188	&	5065.9192	&	-0.0004	\\
$10\nu_2$	&	5151.3969	&	5151.3974	&	-0.0005	\\
\hline
\hline
\end{tabular}
\label{AMES-1B-agree-even}
\end{table}

\section{Line list calculations}

Calculations were performed on the High Performance Computing Service
Darwin cluster, located in Cambridge, UK. Each job from DVR3DRJZ,
ROTLEV3B and DIPOLE3 is submitted to a single computing node
consisting of two 2.60 GHz 8-core Intel Sandy Bridge E5-2670
processors, therefore making use of a total of 16 CPUs each through
OpenMP parallelisation of the various BLAS routines in each module. A
maximum of 36 hours and 64 Gb of RAM are available for each
calculation on a node.  The DVR3DRJZ runs generally did not require
more than 2 hours of wall clock time. The most computationally
demanding parts of the line list calculation are in ROTLEV3B for the
diagonalisation of the Hamiltonian matrices, where wall clock time
increases rapidly with increasing $J$.  Matrix diagonalisation in all
cases was performed using the LAPACK routine DSYEV
\citep{99AnBaBi.method}.

The calculations considered all levels with $J \leq$ 165 and energies
below 15 000 \cm. This gave a total of 3~255~954 energy levels.
Einstein-A coefficients were computed for all allowed transitions
linking any energy level below 8000 \cm\ with any level below 15~000
\cm.  The parameters $J \leq 165$ and $E_{\rm low} = 8000$~\cm\
determine the upper temperature for which the line list is complete;
the upper energy cut-off of 15 000 \cm\ means that this line list is
complete for all transitions longwards of 1.25 $\mu$m. In practice,
the rotation-vibration spectrum of SO$_2$ is very weak at wavelengths
shorter than this and can therefore safely be neglected.  In
the line list, known as ExoAmes, contains a total of 1.3
billion Einstein-A coefficients.

\begin{figure}
\includegraphics[width=\columnwidth]{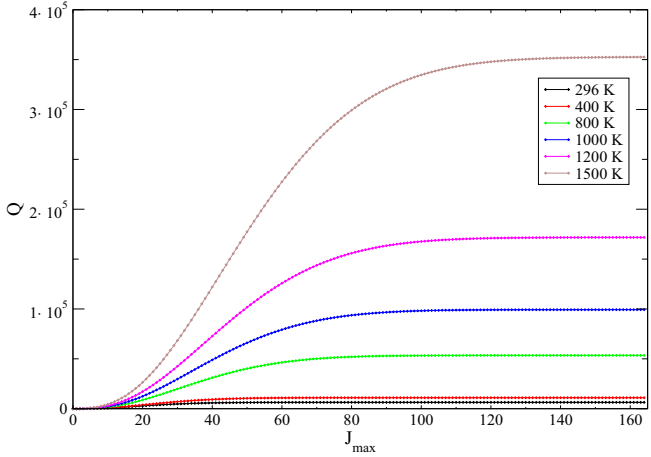}
\caption{Convergence of partition function at different temperatures as a function of $J_{\rm max}$.
The partition function increases monotonically with temperature.}
\label{so2-pfunc-jConv}
\end{figure}

The partition function can be used to assess the completeness of the
line list as a function of temperature, $T$ \citep{jt181}.
The value of the partition function at $T$ = 296 K, computed using all
our energy levels is
6337.131. With a cut-off of $J \leq$ 80, as used by \citet{14HuScLe.SO2},
the value for the same temperature is computed as 6336.803,
which is in excellent agreement with their calculated value of
6336.789.
Figure \ref{so2-pfunc-jConv} shows the partition function values as a
function of a $J$ cut-off for a range of $T$. The highest
value of $J$ considered, $J$ = 165, defines the last point where the
lowest energy is less than 8000 \cm, which is used as the maximum
value of lower energy states in DIPOLE3 calculations. As can be seen
from this figure, the partition function is well converged for $J$ =
165 at all temperatures.

\begin{figure}
\includegraphics[width=\columnwidth]{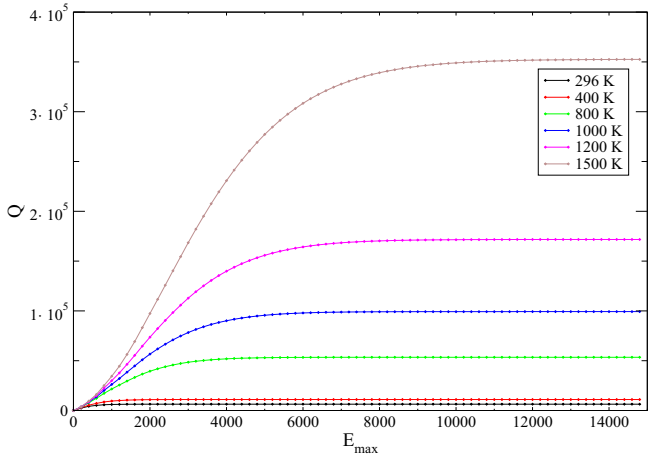}
\caption{Convergence of partition function at different temperatures as a
function of $E_{\rm max}$ (\cm). The partition function increases monotonically with temperature.}
\label{so2-pfunc-eConv}
\end{figure}

The $J$-dependent convergence of $Q$ gives a good indication of the
completeness of the computed energy levels with respect to their
significance at each temperature. However in order to ascertain the
reliability of the line list for increasing temperatures it is more
pertinent to observe the convergence of $Q$ as a function of energy
cut-off; this is illustrated in Figure \ref{so2-pfunc-eConv}.
The importance of this lies in the fact that the computed line list in
the current work only considers transitions from energy levels below
8000 \cm. Since the physical interpretation of an energy level's
contribution to $Q$ is the probability of it's occupancy, the
completeness of the line list can only be guaranteed if all
transitions from states with non-negligible population are computed.
In other words, the line list may only be considered 100 \% complete
if $Q$ is converged when summing over all $E \leq$ 8000 \cm.

\begin{figure}
\includegraphics[width=\columnwidth]{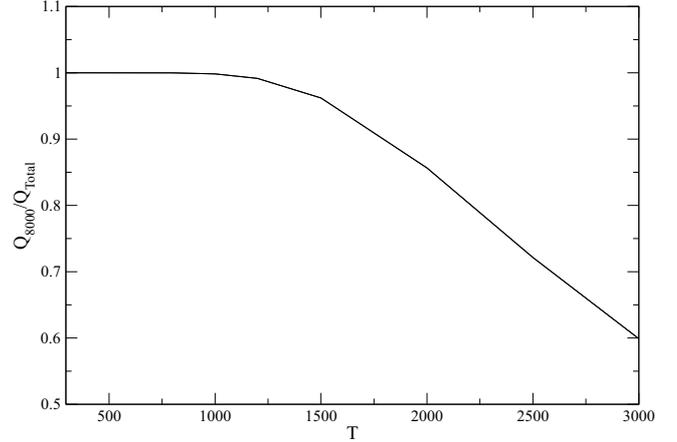}
\caption{Ratios of $Q_{8000}$ to the assumed converged values $Q_{\rm Total}$ as a
function of temperature.}
\label{so2hotPFTemp}
\end{figure}

Figure \ref{so2-pfunc-eConv} shows that, at a cut-off of 8000 \cm, the
partition function is not fully converged for $T$ = 1500 K. Despite
computing all rovibrational levels below 8000 \cm\ ($J \leq$ 165), and
all transitions from these states to states with $E \leq$ 15 000 \cm,
there is still a minor contribution from energies above this cut-off
to the partition sum, corresponding to all values of $J$. However, the
neglected transitions are expected to make only a small additional
contribution to the overall intensity at this temperature.
The completeness of the line list may be quantified by considering the ratio of
the partition function at the 8000 \cm\ cut-off and the total partition
function, $Q_{\rm Total}$, which takes into account all computed energies. Figure
\ref{so2hotPFTemp} shows this ratio as a function of temperature.

For $T \leq$ 1500 K the line list is over 96 \% complete. As can be
seen from Figure \ref{so2hotPFTemp} the level of completeness
decreases with increasing temperature; at 2000 K the ratio falls to 86
\%, and as low as 33 \% for 5000 K. These values assume that
$Q_{\rm Total}$ is equal to the `true' value of the partition function and
tests suggest that for $T= 3000 - 5000$~K, the partition function is
still converged to within 0.1 \%\ when all computed energy levels are
taken into consideration.

\begin{table}
\caption{ Extract from the state file for SO$_2$. The full table is available from
http://cdsarc.u-strasbg.fr/cgi-bin/VizieR?-source=J/MNRAS/xxx/yy.}
\begin{tabular}{rrrrrrrrrrr}
\hline\hline
$i$ & $\tilde{E}$& $g$ & $J$ & $p$ & $\nu_1$ & $\nu_2$ & $\nu_3$ & $K_a$ & $K_c$ \\
\hline
1 & 0.000000 & 1 & 0 & 0 & 0 & 0 & 0 & 0 & 0 \\
2 & 517.872609 & 1 & 0 & 0 & 0 & 1 & 0 & 0 & 0 \\
3 & 1035.118794 & 1 & 0 & 0 & 0 & 2 & 0 & 0 & 0 \\
4 & 1151.714304 & 1 & 0 & 0 & 1 & 0 & 0 & 0 & 0 \\
5 & 1551.759779 & 1 & 0 & 0 & 0 & 3 & 0 & 0 & 0 \\
6 & 1666.328818 & 1 & 0 & 0 & 1 & 1 & 0 & 0 & 0 \\
7 & 2067.808741 & 1 & 0 & 0 & 0 & 4 & 0 & 0 & 0 \\
8 & 2180.319086 & 1 & 0 & 0 & 1 & 2 & 0 & 0 & 0 \\
9 & 2295.815835 & 1 & 0 & 0 & 2 & 0 & 0 & 0 & 0 \\
10 & 2583.270841 & 1 & 0 & 0 & 0 & 5 & 0 & 0 & 0 \\
11 & 2693.705600 & 1 & 0 & 0 & 0 & 4 & 0 & 0 & 0 \\
12 & 2713.393783 & 1 & 0 & 0 & 0 & 0 & 2 & 0 & 0 \\
13 & 2807.174418 & 1 & 0 & 0 & 2 & 1 & 0 & 0 & 0 \\
14 & 3098.143224 & 1 & 0 & 0 & 0 & 6 & 0 & 0 & 0 \\
15 & 3206.501197 & 1 & 0 & 0 & 0 & 5 & 0 & 0 & 0 \\
16 & 3222.952550 & 1 & 0 & 0 & 0 & 1 & 2 & 0 & 0 \\
17 & 3317.908237 & 1 & 0 & 0 & 1 & 3 & 0 & 0 & 0 \\
18 & 3432.272904 & 1 & 0 & 0 & 3 & 0 & 0 & 0 & 0 \\
19 & 3612.415017 & 1 & 0 & 0 & 0 & 7 & 0 & 0 & 0 \\
20 & 3718.711074 & 1 & 0 & 0 & 0 & 6 & 0 & 0 & 0 \\
\hline\hline
\end{tabular}
\label{tab:states}

\mbox{}\\
{\flushleft
$i$:   State counting number.     \\
$\tilde{E}$: State energy in \cm. \\
$g$: State degeneracy.\\
$J$: Total angular momentum            \\
$p$:   Total parity given by $(-1)^{J+p}$. \\
$\nu_1$:   Symmetric stretch quantum number. \\
$\nu_2$:   Bending quantum number. \\
$\nu_3$:   Asymmetric stretch quantum number. \\
$K_a$: Asymmetric top quantum number.\\
$K_c$:  Asymmetric top quantum number.\\
}

\end{table}

Table~\ref{tab:states} gives a portion of the SO$_2$ states file. As
DVR3D does not provide approximate quantum numbers, $K_a$, $K_c$ and
the vibrational labels $\nu_1$, $\nu_2$ and $\nu_3$, these have been
taken from the calculations of \citet{14HuScLe.SO2}, where possible, by
matching $J$, parity and energy; these quantum numbers
are approximate and may be updated in future as better estimates become
available.  Table~\ref{tab:trans} gives a
portion of the transitions file.  This file contains 1.3 billion
transitions and has been split into smaller files for ease of
downloading.

\begin{table}
\caption{Extract from the transitions file for SO$_2$.
The full table is available from
http://cdsarc.u-strasbg.fr/cgi-bin/VizieR?-source=J/MNRAS/xxx/yy.}
\begin{tabular}{rrr}
\hline\hline
$f$ & $i$ & A \\
\hline
679 & 63 & 1.9408E-13 \\
36 & 632 & 5.6747E-13 \\
42 & 643 & 1.7869E-11 \\
635 & 38 & 1.1554E-11 \\
54 & 662 & 3.6097E-11 \\
646 & 44 & 1.9333E-08 \\
660 & 52 & 2.5948E-08 \\
738 & 98 & 3.4273E-06 \\
688 & 69 & 3.4316E-06 \\
47 & 650 & 1.4537E-11 \\
648 & 45 & 3.4352E-06 \\
711 & 82 & 3.5730E-06 \\
665 & 55 & 3.5751E-06 \\
716 & 85 & 3.4635E-06 \\
670 & 58 & 3.4664E-06 \\
635 & 37 & 3.4690E-06 \\
611 & 23 & 3.4701E-06 \\
595 & 12 & 3.4709E-06 \\
734 & 95 & 3.7253E-06 \\
684 & 66 & 3.7257E-06 \\
\hline\hline
\end{tabular}
\label{tab:trans}

\noindent
 $f$: Upper  state counting number;
$i$:  Lower  state counting number; $A_{fi}$:  Einstein-A
coefficient in s$^{-1}$.

\end{table}

\section{Experiments}

\begin{figure}
\includegraphics[width=\columnwidth]{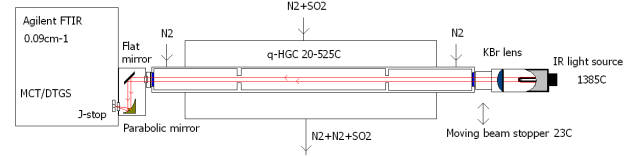}
\caption{Optical set up used in the SO$_2$ infrared absorption measurements.}
\label{f:expt}
\end{figure}

Experiments were performed at the Technical University of Denmark (DTU).
Absorbance measurements for SO$_2$ were performed for temperatures up to
500~C using a quartz high-temperature gas flow cell (q-HGC). This
cell is described in details by \citet{13GrFaNi.SO2} and has recently
been used for measurements of hot NH$_3$ \citep{jt616},
sulphur-containing gases \citep{15GrFaCl.H2S} and some PAH compounds
\citep{15GrSoFa.PAH}.  The optical set-up is shown in
Fig.~\ref{f:expt}. The set-up includes a high-resolution Fourier
transform infraread (FTIR) spectrometer (Agilent 660 with a linearized
MCT and DTGS detectors), the q-HGC and a light-source (Hawkeye,
IR-Si217, 1385C) with a KBr plano-convex lens. The light source is
placed in the focus of the KBr lens. The FTIR and sections between the
FTIR/q-GHC and q-HGC/IR light source were purged using
CO$_2$/H$_2$O-free air obtained from a purge generator.  Bottles with
premixed gas mixture, N$_2$ + SO$_2$ (5000~ppm) (Strandm{\o}llen) and
N$_2$ (99.998\%) (AGA) were used for reference and SO$_2$ absorbance
measurements. Three calibrated mass-flow controllers (Bronkhorst) were
used to balance flow in the middle (N$_2$+SO$_2$) and the two buffer
(N$_2$) parts on the q-HGC and to make additional dilution of the
SO$_2$ to lower concentrations.

SO$_2$ absorbance measurements were performed at 0.25 -- 0.5 \cm\
nominal spectral resolution and at around atmospheric pressure in the
q-HGC. The experimental SO$_2$ absorption spectra were calculated as
described in Section 3.1 of \citet{jt616}. Spectra were recorded in
the range 500 -- 8000 \cm\ and at temperatures of 25, 200, 300, 400
and 500~C.  However at the low SO$_2$ concentrations used the
absorption spectrum was too weak above 2500 \cm\ to yield useful cross
sections. The weak bands centred at 550 \cm\ and 2400 \cm\ are
observed but use of higher concentrations of SO$_2$ is needed to
improve the signal-to-noise ratio. Here we concentrate on the features
in the 1000 -- 1500 \cm\ region. In this region, experimental
uncertainties in the absorption cross sections of the $\nu_3$ band do
not exceed 0.5~\%. This accuracy is confirmed by comparison of 25~C
SO$_2$ absorption cross sections measureed at DTU with those available
in the PNNL (Pacific Northwest National Laboratory) database
\citep{PNNL}.

\subsection{Results}

\subsection{Comparison with HITRAN}

There are 72~459 lines for \sotwo\  in the HITRAN2012 database
\citep{jt557}, which include rovibrational energies up to and
including $J$ = 99. In order to quantitatively compare energy levels
and absolute intensities a similar approach was adopted to that of
\citet{14HuScLe.SO2}.
In order to compare energy levels, the HITRAN transitions are
transformed into a list of levels labelled by their appropriate upper
and lower state quantum numbers; energies are obtained from the usual
lower energy column, $E\pp$, and upper energies are also obtained via
$E\pp + \nu$. Any duplication from the combination difference method
is removed, and energies are only kept if the HITRAN error code for
line position satisfies the condition ierr $\geq$ 4, ensuring all line
position uncertainties are under 1 $\times$ 10$^{-3}$ \cm. For this
reason, the $\nu_1 + \nu_3$, $\nu_1 + \nu_2 + \nu_3$ and 3$\nu_3$
bands are excluded from the current comparison, as by
\citet{14HuScLe.SO2}. This leaves a total of 13 507 rovibrational
levels across 10 vibrational bands available for the comparison which is given
in Table \ref{DVR-HITRAN-compLev}.

\begin{landscape}
\begin{table}
\caption{Comparisons of rovibrational energy levels between available HITRAN data \citep{jt557} and corresponding data calculated using DVR3D.}
\begin{tabular}{lrrrrrrrrr}
\hline\hline
$\nu_1$ $\nu_2$ $\nu_3$	&	$E_{\rm min}$ / \cm	&	$E_{\rm max}$ / \cm	&	$J_{\rm min}$	&	$J_{\rm max}$	&	$K_a^{\rm min}$	&	$K_a^{\rm max}$	&	No.	&	$\Delta_{\rm max}$	&	$\Delta_{\rm RMS}$	\\
\hline
0 0 0	&	1.908	&	4062.964	&	1	&	99	&	0	&	35	&	2774	&	0.092	&	0.014	\\
0 0 1	&	1362.696	&	4085.476	&	1	&	90	&	0	&	33	&	2023	&	0.092	&	0.019	\\
0 0 2	&	2713.383	&	4436.384	&	0	&	76	&	0	&	23	&	1097	&	0.085	&	0.013	\\
0 1 0	&	517.872	&	3775.703	&	0	&	99	&	0	&	29	&	2287	&	0.084	&	0.016	\\
0 2 0	&	1035.126	&	2296.506	&	0	&	62	&	0	&	20	&	894	&	0.073	&	0.010	\\
0 3 0	&	1553.654	&	2237.936	&	0	&	45	&	0	&	17	&	502	&	0.070	&	0.016	\\
1 0 0	&	1151.713	&	3458.565	&	0	&	88	&	0	&	31	&	1706	&	0.097	&	0.016	\\
1 1 0	&	1666.335	&	3080.042	&	0	&	45	&	0	&	21	&	757	&	0.080	&	0.007	\\
0 1 1	&	1876.432	&	3964.388	&	1	&	70	&	0	&	25	&	1424	&	0.087	&	0.017	\\
1 3 0	&	2955.938	&	3789.613	&	11	&	52	&	11	&	11	&	43	&	0.075	&	0.057	\\
\hline
Total	&	1.908	&	4436.384	&	0	&	99	&	0	&	35	&	13507	&	0.097	&	0.016	\\
\hline
\hline

\end{tabular}
\label{DVR-HITRAN-compLev}
\end{table}

\begin{table}
\caption{Statistical summary of comparisons between 13 HITRAN bands and
corresponding bands produced in the current work. Transition frequencies $\nu$
are given in \cm\ and intensities are in cm molecule$^{-1}$.}
\centering
\begin{tabular}{lrrrrrrrrrrrrrr}
\hline\hline
Band	&	$J_f^{\rm max}$	&	$K_a^{\rm max}$	&	$\nu_{\rm min}$	&	$\nu_{\rm max}$	&	No.	&	$\Delta \nu_{\rm max}$	&	$\Delta \nu_{\rm AVG}$	&	$\sigma(\Delta \nu)$	&	Sum I$_{\rm HITRAN}$	&	Sum I$_{\rm ExoAmes}$	&	$\delta$I$_{min}$	&	$\delta$I$_{\rm max}$	&	$\delta$I$_{\rm AVG}$	&	$\sigma$($\delta$I)	\\

\hline
000 $\leftarrow$ 000	&	99	&	35	&	0.017	&	265.860	&	13725	&	0.048	&	0.004	&	0.007	&	2.21$\times$ 10$^{-18}$	&	2.39$\times$ 10$^{-18}$	&	-5.3\%	&	71.0\%	&	14.8\%	&	11.0\%	\\
010 $\leftarrow$ 010	&	99	&	29	&	0.029	&	201.901	&	9215	&	0.041	&	0.004	&	0.006	&	1.78$\times$ 10$^{-19}$	&	1.93$\times$ 10$^{-19}$	&	0.6\%	&	40.6\%	&	10.6\%	&	5.8\%	\\
010 $\leftarrow$ 000	&	70	&	26	&	436.589	&	645.556	&	5914	&	0.030	&	0.006	&	0.005	&	3.71$\times$ 10$^{-18}$	&	3.84$\times$ 10$^{-18}$	&	-48.7\%	&	38.9\%	&	2.5\%	&	18.8\%	\\
020 $\leftarrow$ 010	&	62	&	21	&	446.390	&	622.055	&	3727	&	0.024	&	0.005	&	0.004	&	5.77$\times$ 10$^{-19}$	&	5.94$\times$ 10$^{-19}$	&	-38.7\%	&	34.0\%	&	2.6\%	&	16.1\%	\\
030 $\leftarrow$ 020	&       46	&	17	&	463.097	&	598.267	&	1532	&	0.028	&	0.015	&	0.007	&	5.59$\times$ 10$^{-20}$	&	5.72$\times$ 10$^{-20}$	&	-29.5\%	&	26.0\%	&	2.0\%	&	12.6\%	\\
100 $\leftarrow$ 000	&	88	&	32	&	1030.973	&	1273.175	&	8291	&	0.052	&	0.003	&	0.004	&	3.32$\times$ 10$^{-18}$	&	3.63$\times$ 10$^{-18}$	&	-42.3\%	&	34.6\%	&	6.4\%	&	11.2\%	\\
110 $\leftarrow$ 010	&	45	&	22	&	1047.859	&	1243.820	&	4043	&	0.021	&	0.005	&	0.004	&	2.51$\times$ 10$^{-19}$	&	2.74$\times$ 10$^{-19}$	&	-99.3\%	&	27.4\%	&	6.0\%	&	18.1\%	\\
001 $\leftarrow$ 000	&	90	&	32	&	1294.334	&	1409.983	&	5686	&	0.041	&	0.006	&	0.004	&	2.57$\times$ 10$^{-17}$	&	2.79$\times$ 10$^{-17}$	&	-47.5\%	&	14.1\%	&	1.2\%	&	7.6\%	\\
011 $\leftarrow$ 010	&	71	&	25	&	1302.056	&	1397.007	&	3948	&	0.034	&	0.014	&	0.006	&	2.02$\times$ 10$^{-18}$	&	2.22$\times$ 10$^{-18}$	&	-15.1\%	&	36.0\%	&	5.4\%	&	6.7\%	\\
101 $\leftarrow$ 000	&	82	&	24	&	2433.192	&	2533.195	&	4034	&	0.110	&	0.023	&	0.011	&	5.39$\times$ 10$^{-19}$	&	5.34$\times$ 10$^{-19}$	&	-41.7\%	&	27.2\%	&	-5.3\%	&	5.2\%	\\
111 $\leftarrow$ 010	&	61	&	21	&	2441.079	&	2521.117	&	2733	&	0.145	&	0.011	&	0.013	&	4.24$\times$ 10$^{-20}$	&	4.25$\times$ 10$^{-20}$	&	-30.5\%	&	4.2\%	&	-2.5\%	&	4.1\%	\\
002 $\leftarrow$ 000	&	76	&	24$^{\star}$	&	2599.080	&	2775.076	&	4327	&	0.033	&	0.011	&	0.006	&	3.77$\times$ 10$^{-21}$	&	3.51$\times$ 10$^{-21}$	&	-97.9\%	&	57.9\%	&	-10.8\%	&	13.4\%	\\
003 $\leftarrow$ 000	&	77	&	25	&	3985.185	&	4092.948	&	3655	&	0.122	&	0.031	&	0.030	&	1.55$\times$ 10$^{-21}$	&	1.33$\times$ 10$^{-21}$	&	-94.3\%	&	21.3\%	&	-31.3\%	&	21.1\%	\\
\hline
\hline

\end{tabular}

\noindent
$^{\star}$$K_a$ = 11 excluded.

\label{DVR-HITRAN-compInt}
\end{table}
\end{landscape}

Unsurprisingly, the agreement with the corresponding comparison by
\citet{14HuScLe.SO2} is fairly consistent. There are some minor
deviations in $\Delta_{\rm max}$, though the values of $\Delta_{\rm RMS}$ are
comparable.  These deviations are
largely determined by the use of the Ames-1B PES in the DVR3D calculations.

HITRAN band
positions and intensities are compared to the data produced in the
current work, again in a similar fashion to \citet{14HuScLe.SO2}; all
13 HITRAN bands are compared (despite three of these being excluded
from their energy level comparisons). In Huang et al.'s comparison all
transitions associated with 2$\nu_3$ and $K_a$ = 11 levels were
excluded due to a resonance of the band with $\nu_1$ + 3$\nu_2$; the
same exclusion has been applied here.

A total of 70 830 transitions are available for comparison here,
taking into account those corresponding to energy levels with $J >$
80. A matching criteria close to that used for energy level matching,
with the addition that the Obs. - Calc. residuals for $\nu$ also
satisfy $\leq$ 0.2 \cm\ was used. The algorithm used is prone to
double-matching, leading to comparisons which may be reasonable in
wavenumber residuals but not in intensity deviations. In these
instances, the intensity comparisons are screened via the symmetric
residual \citep{14HuScLe.SO2} $\delta(I)$\% = 50\% $\times (I_{\rm
  ExoAmes}/I_{\rm HITRAN} - I_{\rm HITRAN}/I_{\rm ExoAmes})$, where
the best match is found where this value is at a minimum. These
criteria have been able to match all available lines, with the
exception of the 001 $-$ 000 band which matches only 5686 out of 5721
lines.  35 lines were excluded from our statistics because  there
appears to be systematic errors in HITRAN for energy levels with $K_a \geq
33$, see \citet{13UlOnGr.SO2} and \citet{14HuScLe.SO2}.  Table
\ref{DVR-HITRAN-compInt} shows a statistical summary of the band
comparisons.

The standard deviations in line position, $\sigma(\Delta \nu)$, and
line intensity, $\sigma$($\delta$I), are in fairly good agreement with
those of Huang \etal\ despite the use of a different PES, and the
inclusion of energies with $J >$ 80. The differences in minimum, maximum, and
average values may be attributed to our inclusion of higher $J$ levels, though the tighter
restriction on the line intensity matching algorithm used in this work
may also contribute.

\subsection{Comparison with High-Temperature Measurements}

\begin{figure}
\includegraphics[width=\columnwidth]{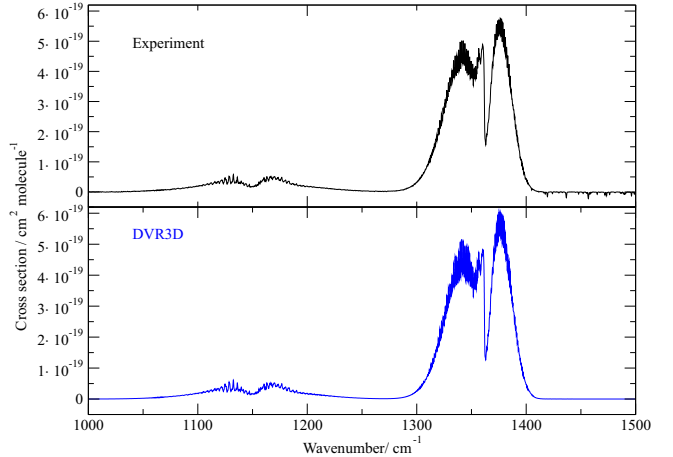}
\caption{Comparison of absorption cross sections obtained at $T$ = 573.15 K for \sotwo\
experimentally (above) and from the hot line
list (below).}
\label{so2-xsec-300c}
\end{figure}

\begin{figure}
\includegraphics[width=\columnwidth]{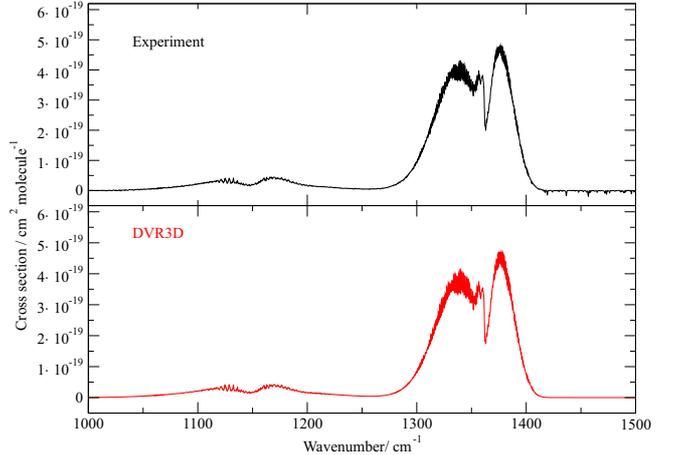}
\caption{Comparison of absorption cross sections obtained at $T$ = 773.15 K for \sotwo\ via
experimentally (above) and from the hot line
list (below).}
\label{so2-xsec-500c}
\end{figure}

Figures \ref{so2-xsec-300c} and \ref{so2-xsec-500c} show the simulated
cross sections for the 1000 $< \nu\ < $ 1500 \cm\ spectral region at
573.15 K (300 C) and 773.15 K (500 C), respectively, convolved with a
Gaussian line shape function with HWHM = 0.25 \cm. These are compared
with experimental cross sections measured at
a resolution of 0.5 \cm. The simulations are calculated using a cross
section code, `ExoCross', developed  work
with the ExoMol line list format \citep{jt548,jt631}, based on the
principles outlined by \citet{jt542}.

This spectral region considered contains both the $\nu_1$ and $\nu_3$
bands, and the intensity features are qualitatively well represented
by the simulated cross sections. For 573.15 K (300 C) the integrated
intensity across the 1000 $< \nu\ < $ 1500 \cm\ spectral region is
calculated as 3.43 $\times$ 10$^{-17}$ cm$^2$ molecule$^{-1}$, which
is about 2\% less than that for the experimental value, measured as
3.50 $\times$ 10$^{-17}$ cm$^2$ molecule$^{-1}$.

For 773.15 K (500 C) the integrated cross section across the same spectral
region is calculated as 3.41 $\times$ 10$^{-17}$ cm$^2$
molecule$^{-1}$, which is roughly 6\% less than that for the
experimental value, 3.62 $\times$ 10$^{-17}$ cm$^2$ molecule$^{-1}$.
This may be attributed to a small discrepancy observed in the P-branch
of the $\nu_3$ band which is not obvious from Figure
\ref{so2-xsec-500c}; the intensity here is slightly lower for the
computed cross sections. Since this disagreement affects a specific
region of the spectrum, it is unlikely wholly due to an error in the
partition sum. The quality of the DMS may also be a contributing
factor, in conjunction with the states involved in these transitions.
Another source may be from the generation of the cross sections
themselves; the line shape function used in constructing the
theoretical cross sections is Gaussian, and therefore only considers
thermal (Doppler) broadening, as opposed to a combination of thermal
and pressure broadening (Voigt line shape). It is possible that
neglecting the (unknown) pressure-broadening contribution in the line shape
convolution is the source of this disagreement. Regardless of this
discrepancy, use of a Voigt profile would considerably improve the
overall quality of computed cross sections.

\subsection{Cross Sections}

\begin{figure}
\includegraphics[width=\columnwidth]{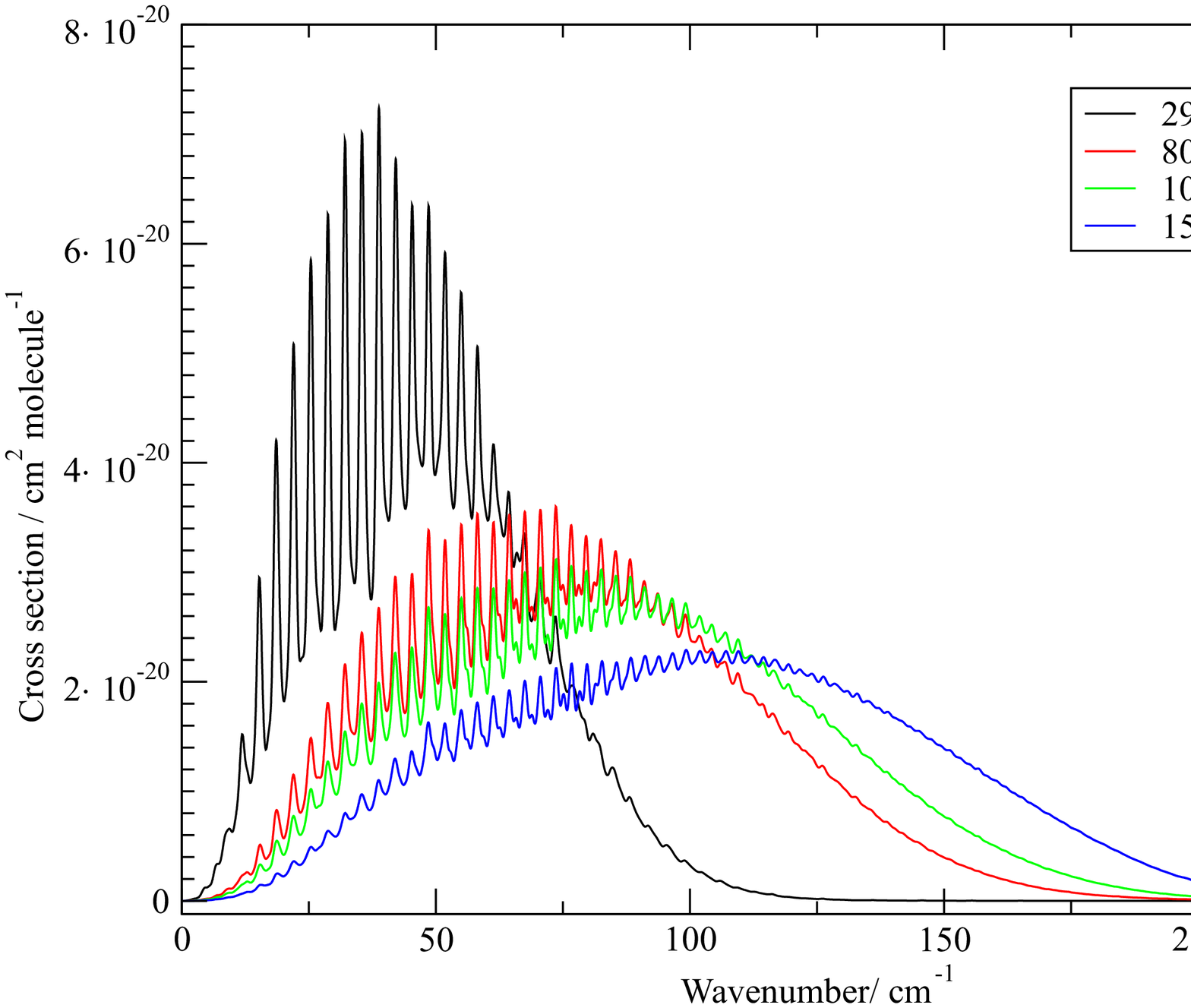}
\caption{Temperature-dependent absorption cross sections for the rotational band of \sotwo.
The maximum peak heights decrease monotonically with temperature.}
\label{so2-xsec-rot}
\end{figure}

\begin{figure}
\includegraphics[width=\columnwidth]{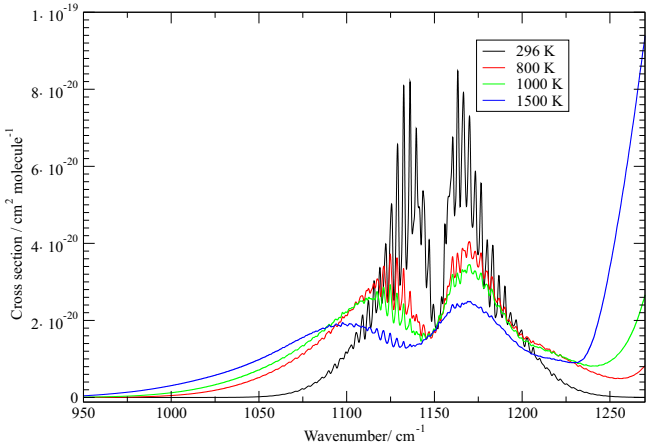}
\caption{Temperature-dependent absorption cross sections for the $\nu_1$ band of \sotwo. The contribution to the intensity beyond 1225 \cm\ is due to the $\nu_3$ band. The maximum peak heights decrease monotonically with temperature.}
\label{so2-xsec-nu1}
\end{figure}

\begin{figure}
\includegraphics[width=\columnwidth]{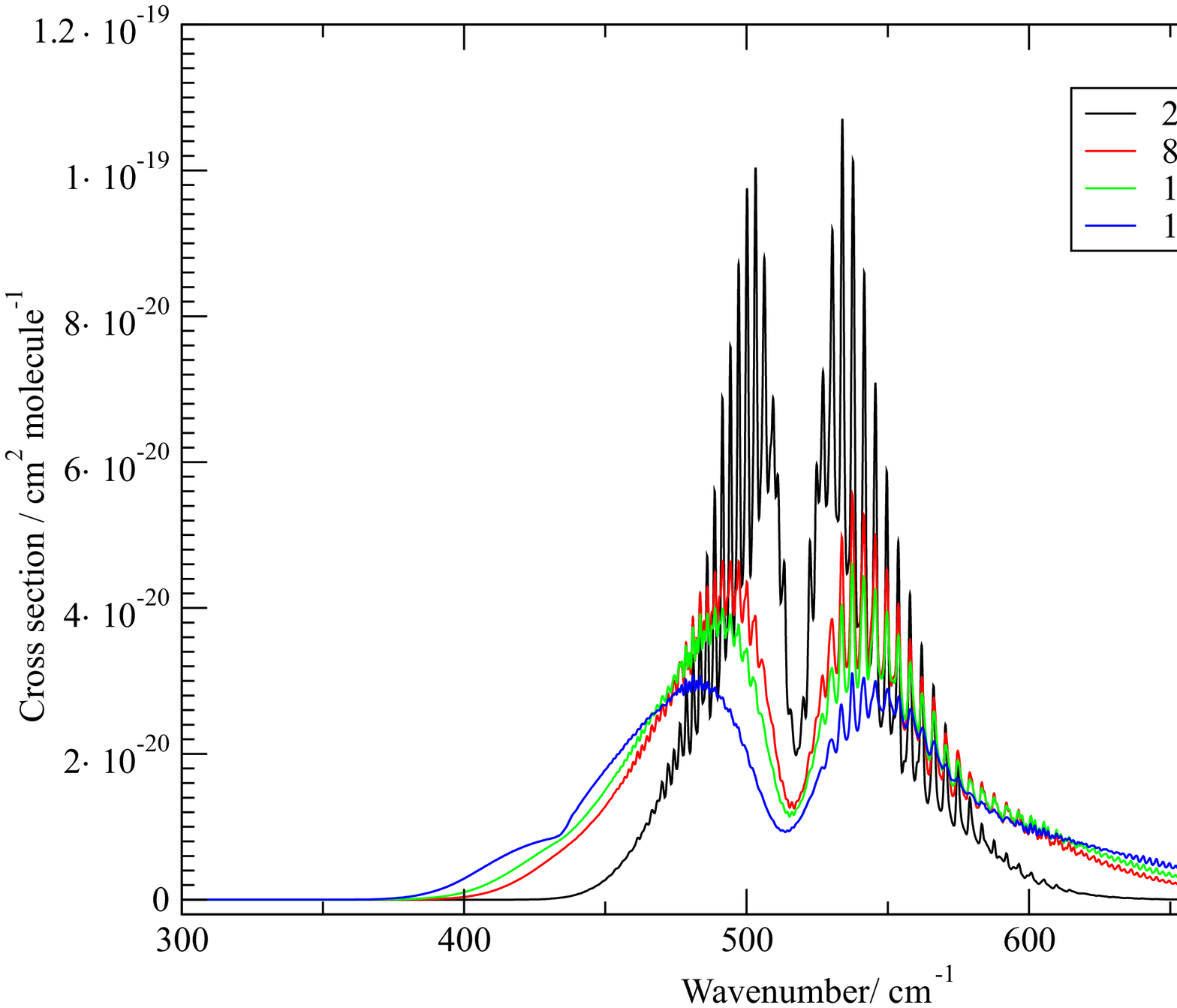}
\caption{Temperature-dependent absorption cross sections for the $\nu_2$ band of \sotwo.
The maximum peak heights decrease monotonically with temperature.}
\label{so2-xsec-nu2}
\end{figure}

\begin{figure}
\includegraphics[width=\columnwidth]{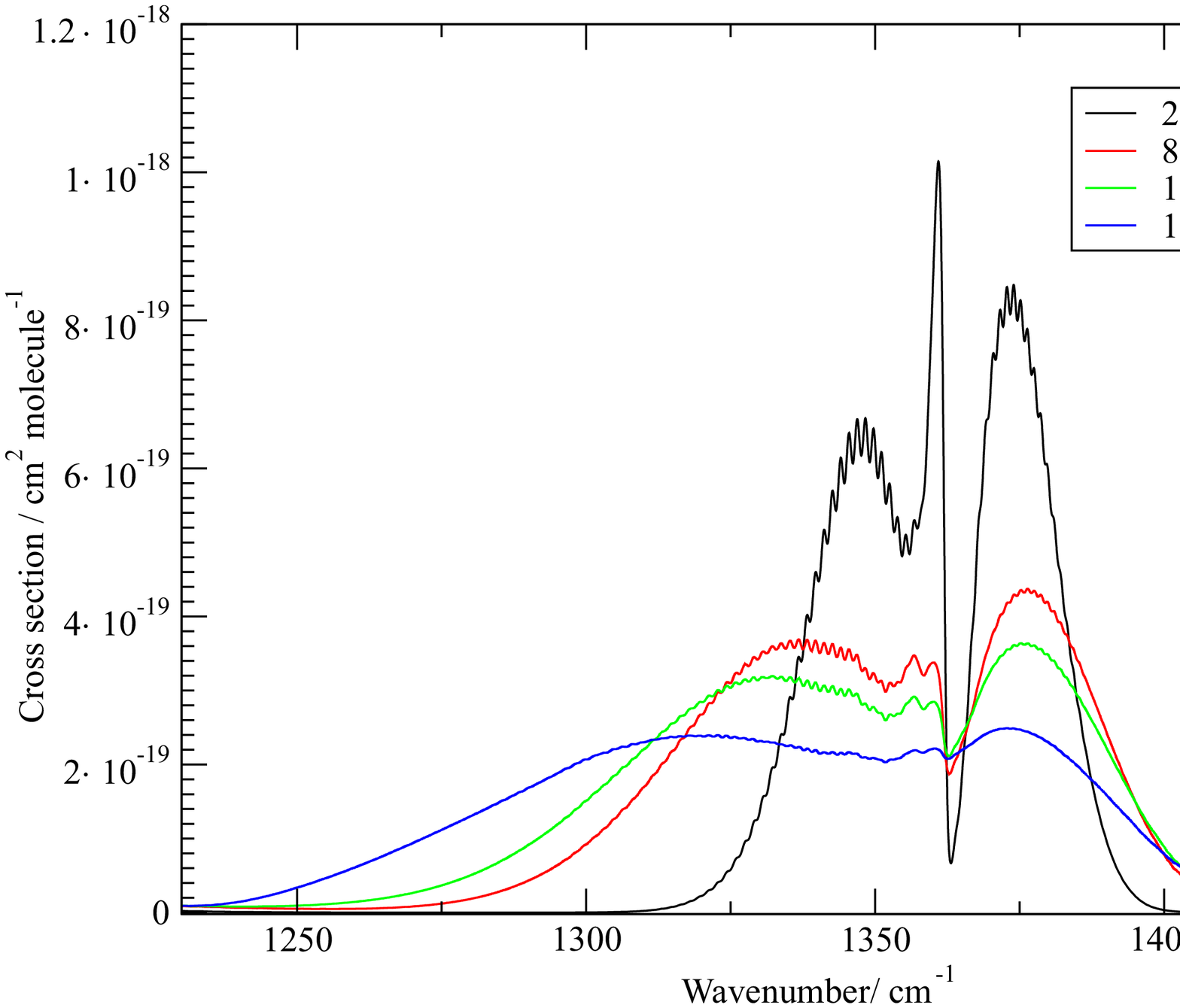}
\caption{Temperature-dependent absorption cross sections for the $\nu_3$ band of \sotwo.
The maximum peak heights decrease monotonically with temperature.}
\label{so2-xsec-nu3}
\end{figure}

Figures \ref{so2-xsec-rot} - \ref{so2-xsec-nu3} display
temperature-dependent calculated cross sections for the rotational and
two fundamental bands of \sotwo. All simulations are produced using the
hot line list convolved with a Gaussian line shape function with HWHM
= 0.5 \cm.

\begin{figure}
\includegraphics[width=\columnwidth]{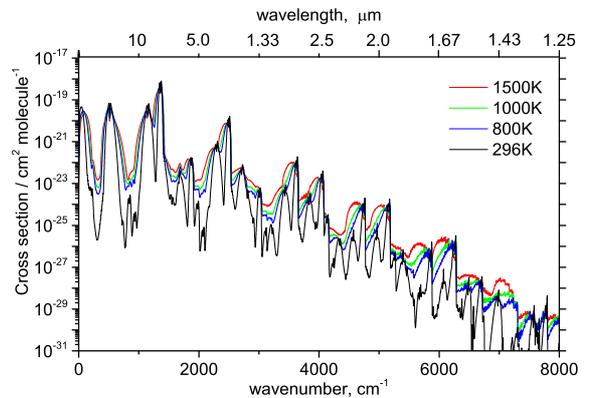}
\caption{Temperature-dependent absorption cross sections for the entire $\lambda > 1.25$~$\mu$m or 0 $< \nu\ < $ 8000 \cm\ region of \sotwo. The dips in the cross sections are progressively smoothed out with increading temperature.}
\label{so2-xsec-full}
\end{figure}

Figure \ref{so2-xsec-full} shows an overview plot of the spectrum for
0 $< \nu\ < $ 8000 \cm ($\lambda > 1.25$~$\mu$m), highlighting the temperature-dependence of the
cross section intensities. Again, this simulation is produced using
the hot line list convolved with a Gaussian line shape function with
HWHM = 2.0 \cm.

\section{Conclusion}

A new hot line list for SO$_2$, called ExoAmes, has been computed
containing 1.3 billion transitions. The line list is divided into an
energy file and a transitions file. This is done using the standard
ExoMol format \citep{jt548} based on the method originally developed
for the BT2 line list by \citet{jt378}.  The full line list can be
downloaded from the CDS, via
\url{ftp://cdsarc.u-strasbg.fr/pub/cats/J/MNRAS/xxx/yy}, or
\url{http://cdsarc.u-strasbg.fr/viz-bin/qcat?J/MNRAS//xxx/yy}, as well
as the exomol website, \url{www.exomol.com}.  The line lists and
partition function together with auxiliary data including the
potential parameters and dipole moment functions, as well as the
absorption spectrum given in cross section format \citep{jt542}, can
all be obtained also from \url{www.exomol.com} as part of the extended
ExoMol database \citep{jt631}.

SO$_2$ is one of three astrophysically-important sulphur oxides.
A room temperature line list for SO$_3$ has already been computed \citep{jt554}
and a hot line list has recently been completed. The results of these
calculations will be compared with recent observations recorded at DTU.
The comparison and the line list will be presented here soon.

Unlike SO$_2$ and SO$_3$, SO is an open shell system with a $^3\Sigma^+$
symmetry electronic ground states which therefore requires special treatment
\citep{15Schwenke.diatom}. A line list for this system will soon be computed
with the program {\sc Duo} \citep{jt609} which has been newly-developed
for treating precisely this sort of problem.

\section*{Acknowledgements}

This work was supported by Energinet.dk project 2010-1-10442 \lq\lq
Sulfur trioxide measurement technique for energy systems'' and the ERC
under the Advanced Investigator Project 267219. It made use of the
DiRAC@Darwin HPC cluster which is part of the DiRAC UK HPC facility
for particle physics, astrophysics and cosmology and is supported by
STFC and BIS.  XH, DWS, and TJL gratefully acknowledge funding support
from the NASA Grant 12-APRA12-0107.  XH also acknowledges support
from the NASA/SETI Institute Cooperative Agreement NNX15AF45A.

\bibliographystyle{mnras}
\label{lastpage}

\end{document}